 \documentclass[final,3p,times,12pt]{elsarticle}
\usepackage{amssymb}
\usepackage{subfigure}          %inclusion of small, `sub,' figures and tables.

 \usepackage{lineno}

\journal{Astroparticle Physics}

\begin{document}

%\modulolinenumbers[1]
%\linenumbers

\begin{frontmatter}



\title{Sensitivity to primary composition and hadronic models from average shape of high energy cosmic ray shower profiles}

 \author[LIP]{S. Andringa}
% \author[LIP]{L. Cazon}
 \author[LIP]{R. Concei\c{c}\~{a}o\corref{cor1}}
 \ead{ruben@lip.pt}
 \cortext[cor1]{Corresponding author}
 \author[LIP]{F. Diogo}
 \author[LIP,IST]{M. Pimenta}
 \address[LIP]{LIP, Av. Elias Garcia, 14-1, 1000-149 Lisboa, Portugal}
 \address[IST]{Departamento de F\'{i}sica, IST, Av. Rovisco Pais, 1049-001 Lisboa, Portugal}

\begin{abstract}
The concept of Universal Shower Profile is used to characterize the average behavior of high energy cosmic rays. 
The shape variables contain important information about composition. They are independent of the primary cross-section by construction, but affected by other hadronic parameters, like multiplicity. The two variables give access to the average nuclear mass of the sample and their compatibility serves as a test of hadronic models.
\end{abstract}


\end{frontmatter}

% \linenumbers

\section{Introduction}
\label{sec:Intro}

The development of high energy cosmic ray showers in the atmosphere gives indirect information on the primary particle. 
It is, however, governed by very high energy hadronic interactions for which there is no unique description, only different 
phenomenological models that try to extrapolate the accelerator data obtained at lower energies and different phase space regions. 

In general the maximum number of particles and the depth at which that maximum occurs are used to characterize the showers. 
The first is a measurement of energy, while the the second statistically distinguishes primary particles 
and depends on the cross-section of the first interaction. More information is however hidden in the details of the 
shower longitudinal profile shape \cite{USPV,UnivM,UnivG}.

Electromagnetic (and muonic \cite{USPmu}) longitudinal shower profiles are well described by the Gaisser-Hillas function, 
which can be parametrized in $N'=N/N_{max}$ and $X'=X-X_{max}$ as:
\begin{equation}
N'=\left(1+\frac{RX'}{L}\right)^{R^{-2}} \exp{\left(-\frac{X'}{LR}\right)}
%N'=\exp\Bigl(-\frac{1}{2}\Bigl(\frac{X'}{L}\Bigr)^2\Bigr)\prod_{n=3}^{\infty}{\exp\Bigl(-\frac{R^{n-2}}{n}\Bigl(-\frac{X'}{L}\Bigr)^n\Bigr)}
\end{equation}
With the above parametrization, introduced in \cite{USPV}, the profile can be recognized as a Gaussian, with width {\bf L},  with an asymmetry introduced by non-zero values of {\bf R}. Notice that after translation to the maximum, the variations due to the point of first interaction disappear\footnote{Although changes in the first interaction point affect the atmospheric profile crossed by the air shower, the corresponding effect on the profile shape in gcm$^{-2}$ is negligible when compared to the differences between the several hadronic interaction models used for the shower development, for example.}.

We will start by introducing the Universal Shower Profile (USP), showing that the shape of the energy deposit longitudinal profile is almost universal. Both shape variables, {\bf L} and {\bf R}, tend to different values according to energy, primary particle and hadronic interaction model. We will then study the construction of the USP for a given energy, regardless of the possible mixed compositions. In fact, both parameters are a function of the average nuclear mass, with different energy evolutions. We will find that the different hadronic interaction models define specific possible regions for the two parameters, which can then be used to exclude or constrain present models or help in the construction of new ones. 

\section{Shower Profile Universality}

The universality of the shower longitudinal profiles, when expressed in  $N'\equiv N/N_{max}$ and $X'\equiv X-X_{max}$ was discussed in \cite{USPV}, where it was shown that {\bf L} and {\bf R} could give extra information on the shower properties. Here we concentrate on what can be learned from the average showers. In fact, the stability of the above variables within showers initiated by similar primaries means that by collecting only a limited number of events, important information can already be obtained, with a significant reduction of statistical uncertainties.

The average shower profile is not exactly described by a Gaisser-Hillas and so the values of {\bf L} and {\bf R} can vary slightly according to the $X'$ interval considered in their determination. In order to increase the available data in a real experiment, the total $X'$ interval should be minimized. However, for a good determination of the width, {\bf L}, a minimum range around the maximum is needed, while for {\bf R} the rising tail needs to be included\cite{USPV}. From now on, we will always consider the range of $X' \in [-350,200]$~g/cm$^2$.

\begin{figure}[htp]
\begin{center}
\includegraphics[width=0.45\textwidth]{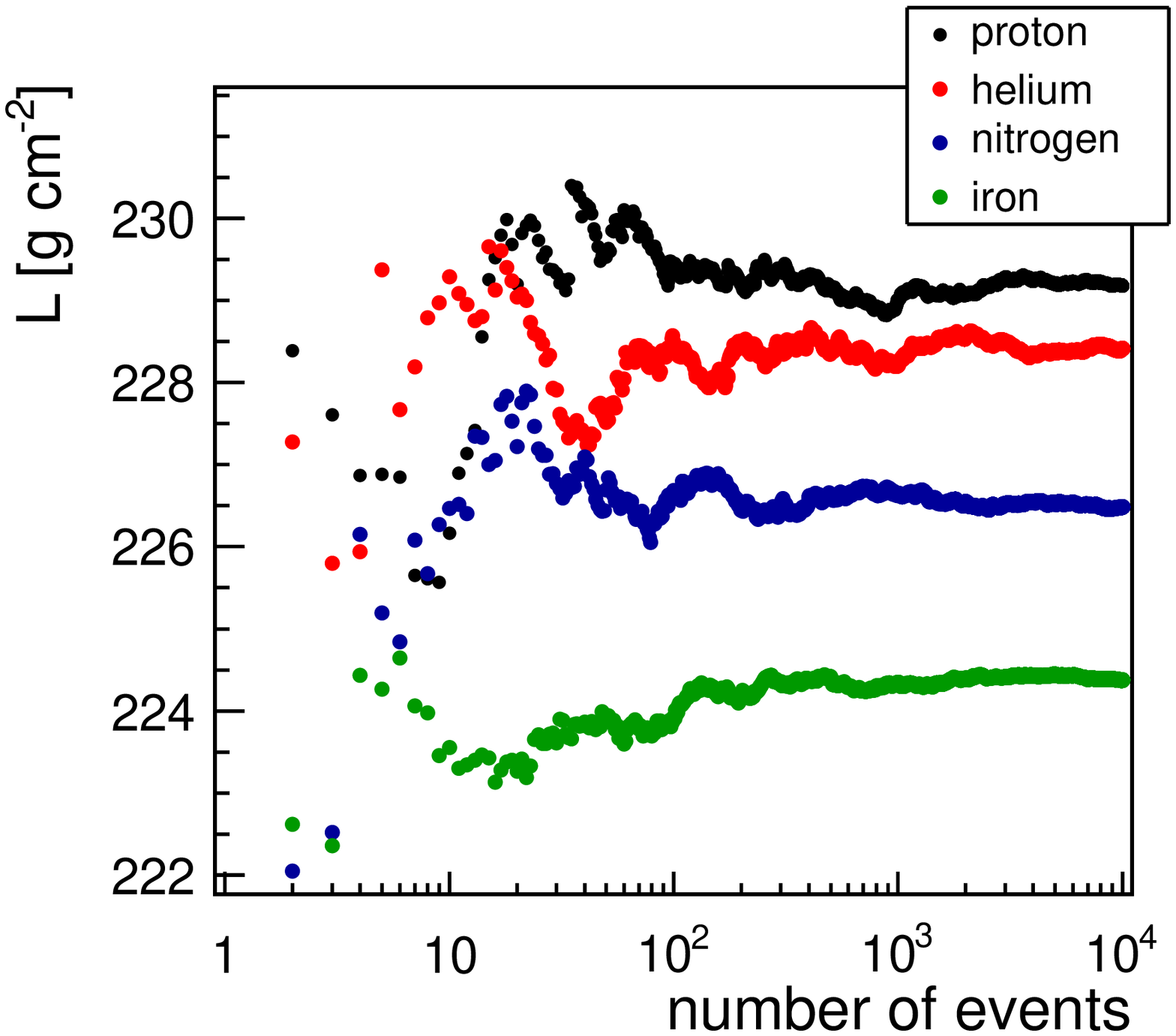}
\includegraphics[width=0.45\textwidth]{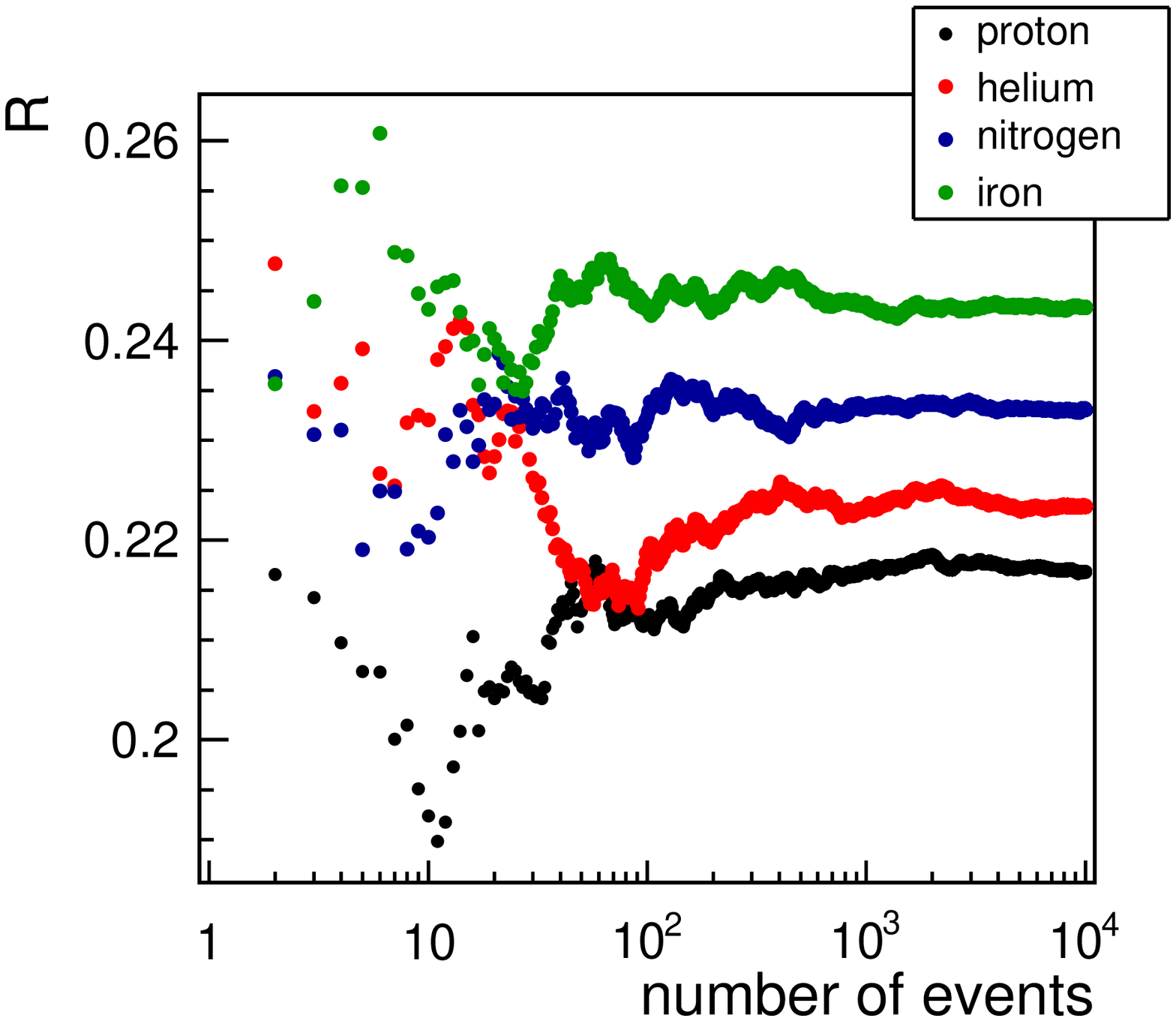}
\caption{Shape parameters {\bf L} (left) and {\bf R} (right) as a function of the number of events used to get the average shower. 
The showers were generated using QGSJet-II as the high energy hadronic interaction model.
Color codes show different primaries at $\log(E/{\rm eV})=19.0$.}
\label{fig:RL}
\end{center}
\end{figure}

\begin{figure}[htp]
\begin{center}
\includegraphics[width=0.45\textwidth]{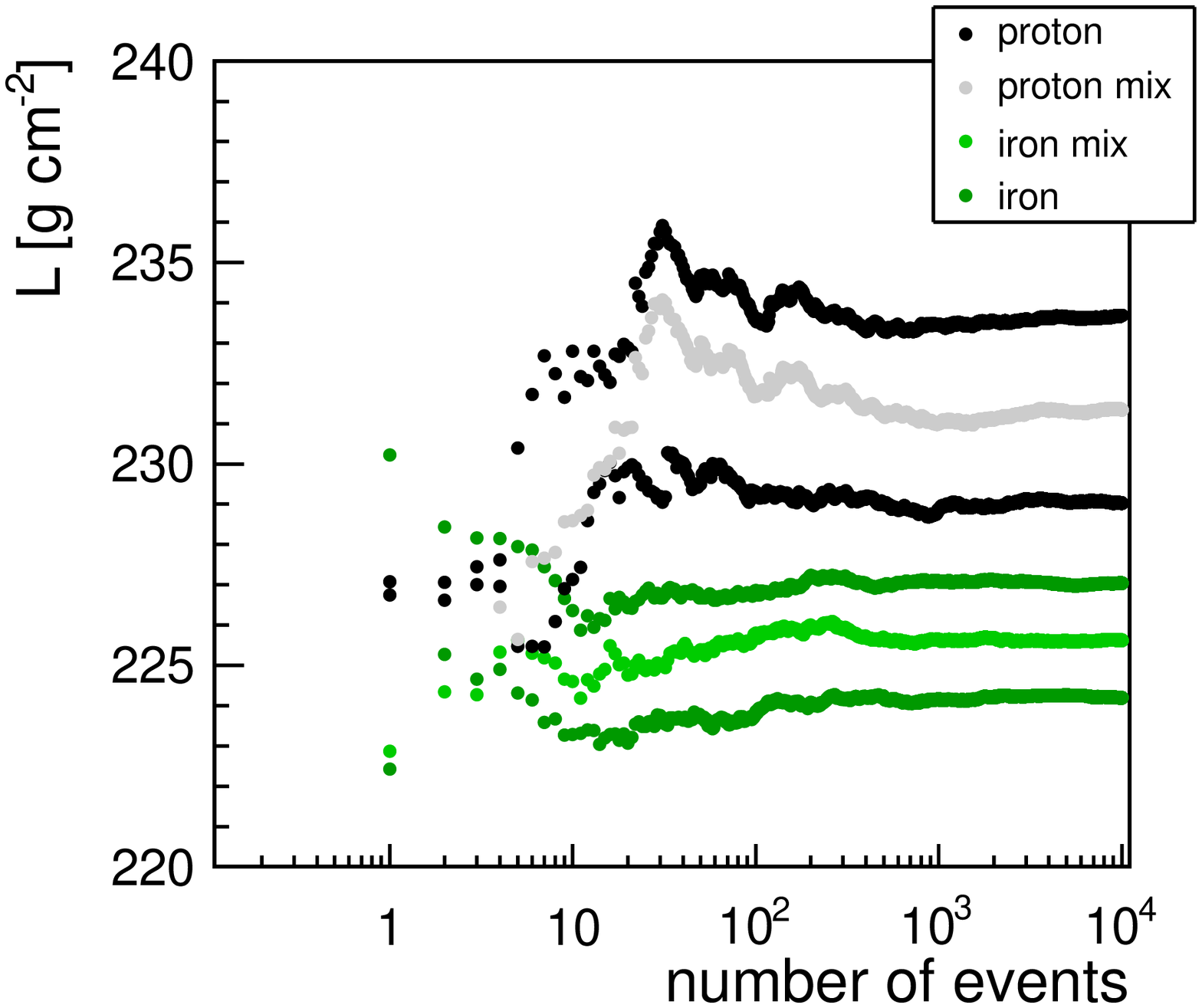}
\includegraphics[width=0.45\textwidth]{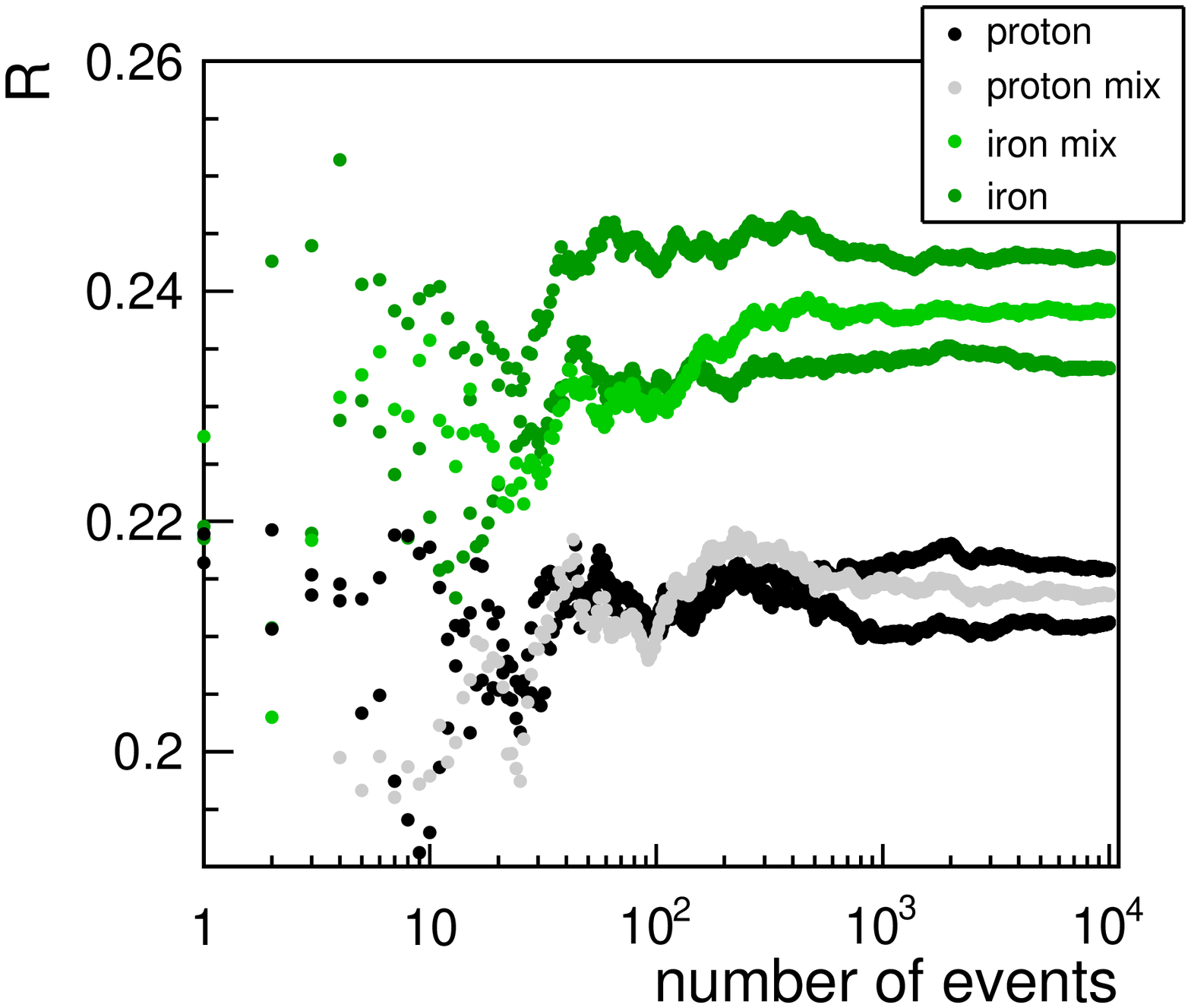}
\caption{Shape parameters {\bf L} (left) and {\bf R} (right) as a function of the number of events used to get the average shower. 
The showers were generated using QGSJet-II as the high energy hadronic interaction model. 
Darker colors show fits to the USP of proton and iron showers at both $\log(E/{\rm eV})=18.5$ and $\log(E/{\rm eV})=19.0$;
light colors correspond to the USP obtained by mixing events generated at two energies evenly.}
\label{fig:RLene}
\end{center}
\end{figure}

Figure \ref{fig:RL} shows the build up of the USP variables, with the collection of events. The energy deposit shower profiles were generated with CONEX\cite{Conex1,Conex2}, using the QGSJet-II.03\cite{QGSII1,QGSII2} model, for different primaries at fixed energy of $\log(E/{\rm eV})=19$. After around one hundred events, the average shape becomes stable, with a good separation for the different primaries, in terms of both variables. 

The USP variables evolve with energy, and figure \ref{fig:RLene} shows the results for proton and iron primaries of different energies. Average values of {\bf L} and {\bf R} are reached when different energies are mixed. In general, data analysis will mix a range of energies, and care must be taken in the interpretation of the results. However, it can be seen that even for a wide interval of $\log(E/{\rm eV}) \in [18.5,19.0]$, light and heavy primaries can still be distinguished.

\section{USP and Mass Composition}

\begin{figure}[htp]
\begin{center}
\includegraphics[width=0.45\textwidth]{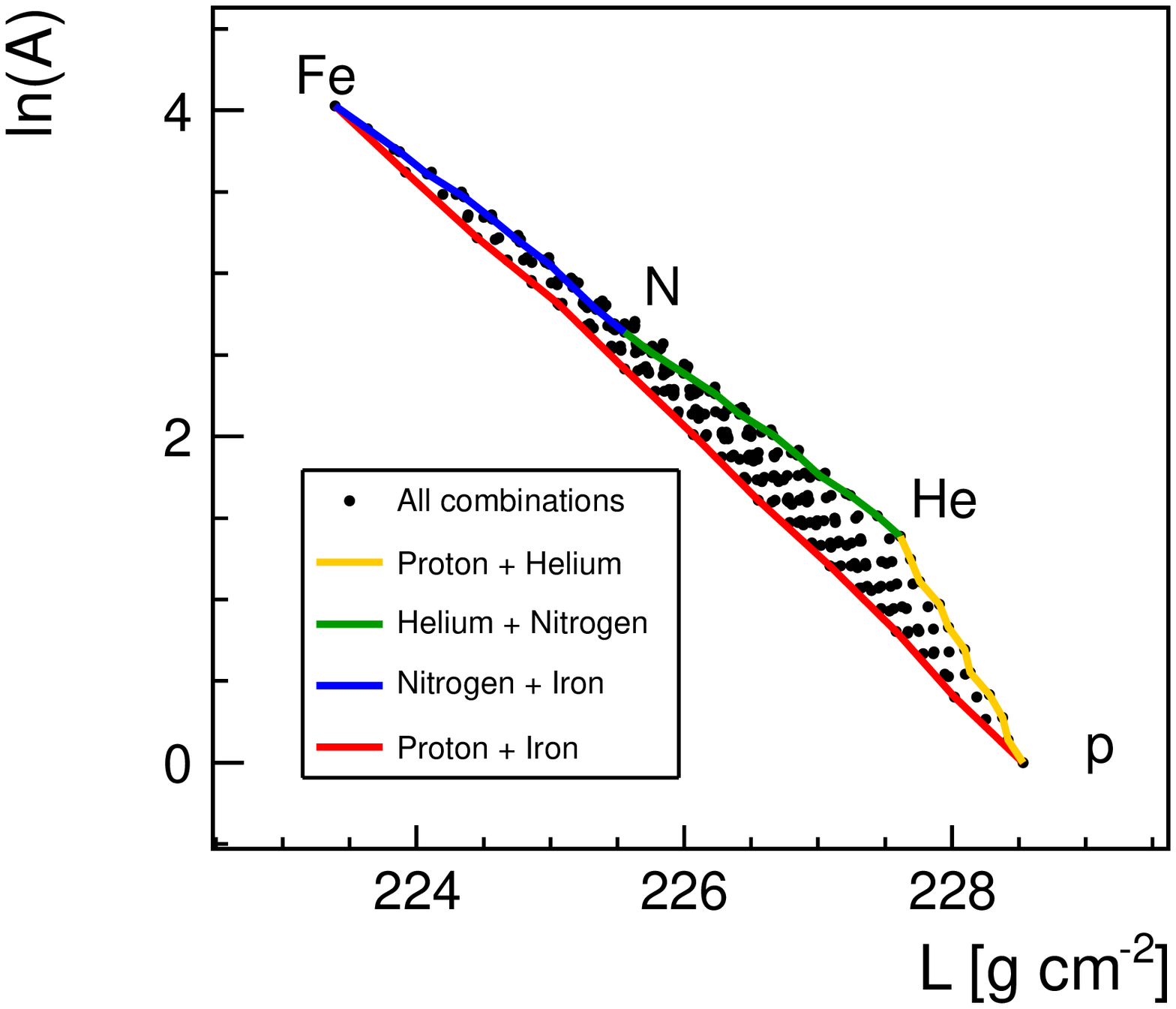}
\includegraphics[width=0.45\textwidth]{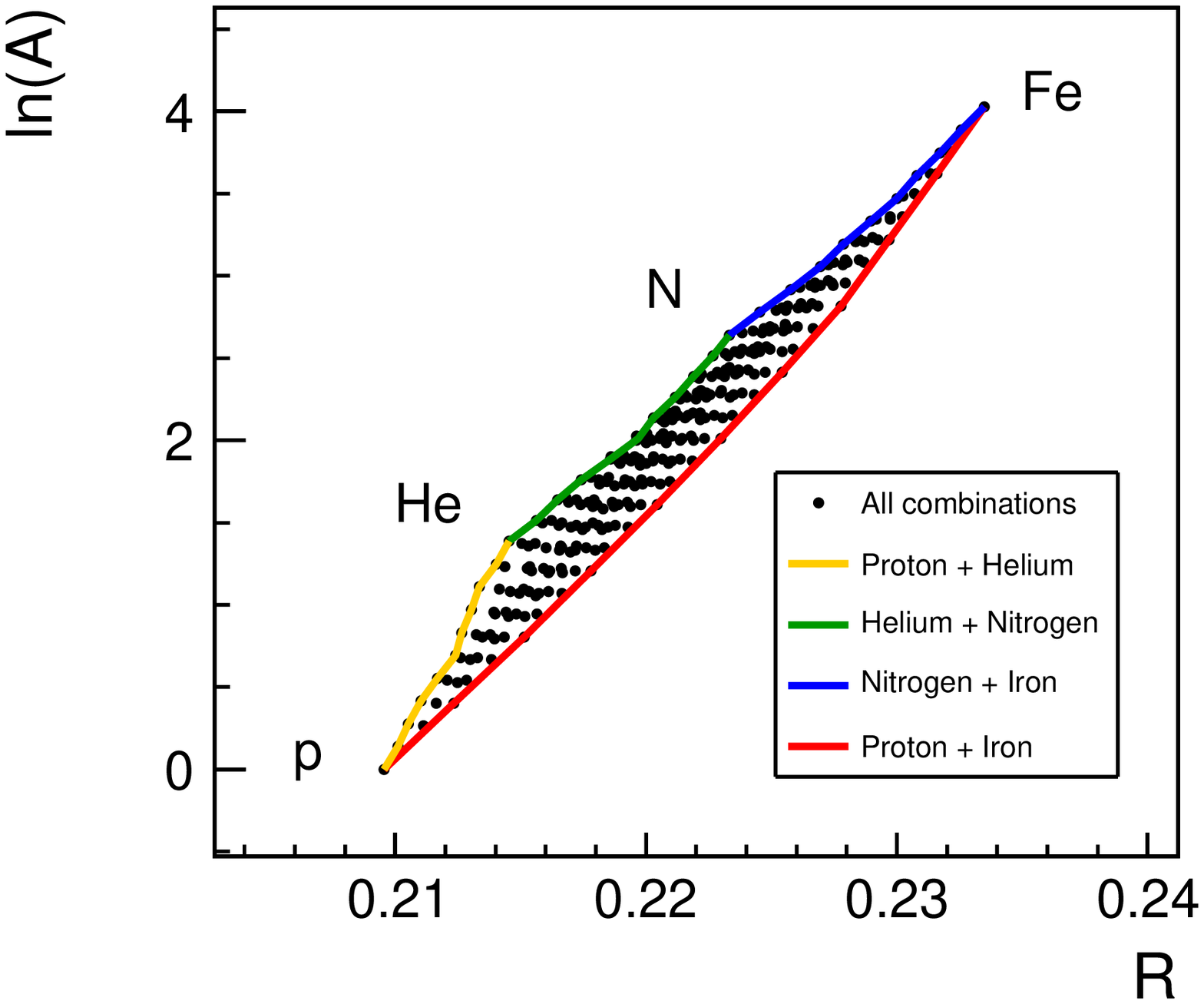}
\caption{Shape parameters {\bf L} (left) and {\bf R} (right) as a function of $\left< ln(A)  \right>$.
The showers were generated using QGSJet-II as the high energy hadronic interaction model. 
Color codes show different combinations of primaries, generated at $\log(E/{\rm eV})=19$.}
\label{fig:RL1D}
\end{center}
\end{figure}

\begin{figure}[h]
\begin{center}
\includegraphics[width=0.85\textwidth]{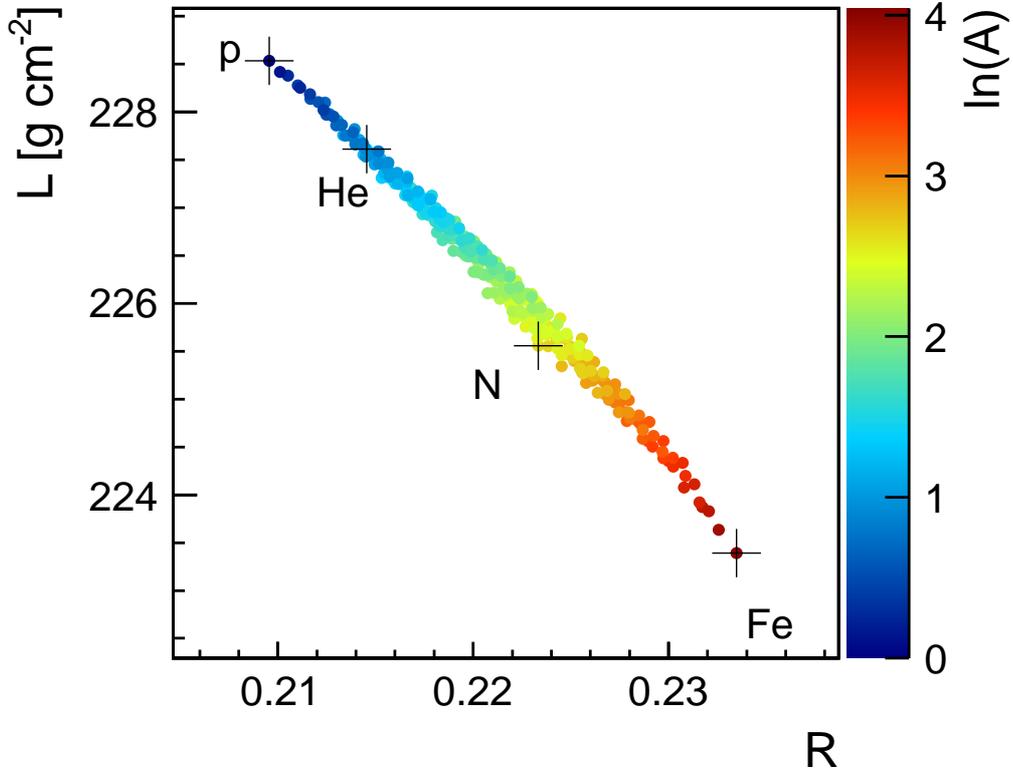}
\caption{Relation of the shape parameters of the universal shower profile with the $\left< ln(A) \right>$. 
The showers were generated using QGSJet-II as the high energy hadronic interaction model.
Color codes show the $\left< ln(A) \right>$ for different combinations of primaries, generated at $\log(E/{\rm eV})=19$,
and crosses mark the results for pure composition samples.}
\label{fig:RL2D}
\end{center}
\end{figure}

The values of {\bf L} and {\bf R} have an almost linear dependence with the logarithm of the nuclear mass, $ln(A)$, as can be seen in fig. \ref{fig:RL1D}. There, in addition to the pure samples, also different mixtures of primaries are presented. Even in these cases, the stability around an average shape is easily achieved.

The values of both parameters give two measurements of $\left< ln(A) \right>$, with small variation according to the mixture of the sample. A pure sample of helium primaries will have an only slightly lower value of {\bf R} and higher value of {\bf L} than a mixture of proton and iron primaries with the same average logarithmic mass. Since the values of both parameters correlate with $\left< ln(A) \right>$, a well defined region can be expected in the {\bf L, R} plane, as shown in fig. \ref{fig:RL2D}. 

Notice also that the same analysis can be extended to include $X_{max}$, bearing in mind that in that case there is an influence of the primary cross-section: a big change of the cross-section (as, for example, suggested in ref.~\cite{xsec}) would then be apparent as a consistent reading of $\left< ln(A) \right>$ between {\bf L} and {\bf R}, with an inconsistency for $\left< X_{max} \right>$. 

\section{Primary composition and hadronic parameters}

The different values of the shape variables {\bf L} and {\bf R} according to primary composition, energy and hadronic interaction models can be used to understand the hadronic parameters governing them. While the new observables are insensitive to the point of first interaction, they can be deeply changed by the most usual processes in cascade development. 

\begin{figure}[htp]
\begin{center}
\includegraphics[width=0.85\textwidth]{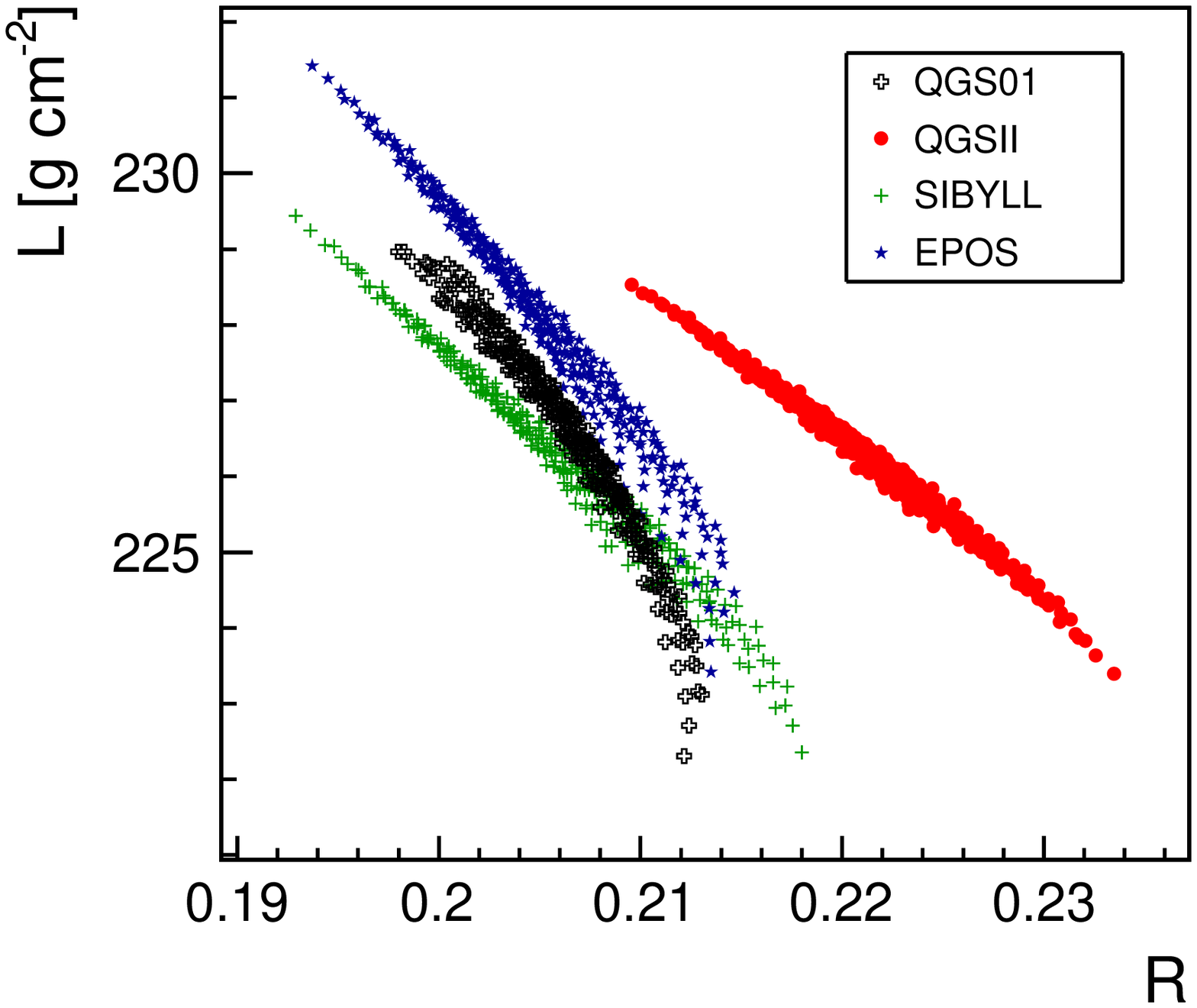}
\caption{Shape parameters {\bf (L, R)} of the universal shower profile , generated at $\log(E/{\rm eV})=19$,
for the different mass composition combinations in several models QGSJet-II.03, SIBYLL2.1, EPOS1.99, QGSJet01c.}
\label{fig:models}
\end{center}
\end{figure}

The correlated evolution of {\bf L} and {\bf R} is characteristic of the hadronic interaction model. Figure \ref{fig:models} shows the comparison of four different high energy hadronic interaction models: QGSJet01c\cite{QGS01}, SIBYLL2.1\cite{SYBILL21} and EPOS1.99\cite{EPOS}, in addition to QGSJet-II.03, for showers at $\log(E/{\rm eV})=19$. The measurement of one point in this plane can be used to check the compatibility of each model with experimental data.

It can be seen, for example, that EPOS has larger {\bf L} values for proton showers than the others; it is even more interesting to note that QGSJet-II spans a range of the {\bf L, R} plane which is quite different from the other three models, namely in what concerns the larger {\bf R} values for heavy nuclei. 
This might reflect the fact that multiplicity is much higher in this model. 
High values of {\bf R} characterize heavy primaries, which also have larger multiplicities in the first interactions, and the same kind of physical characteristics that enable the distinction of primaries is expected to be present between hadronic interaction models.

\begin{figure}[htp]
\begin{center}
\includegraphics[width=0.45\textwidth]{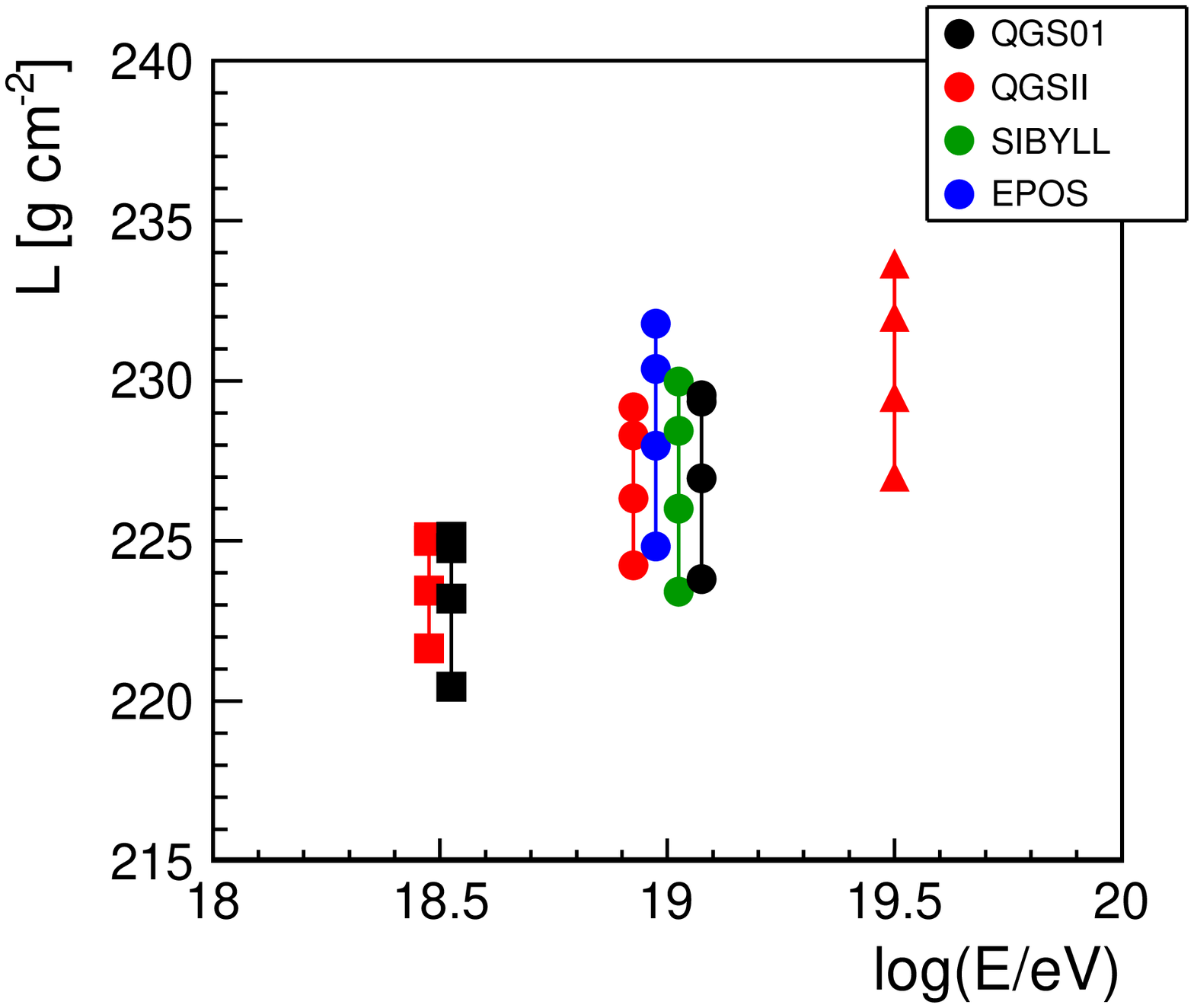}
\includegraphics[width=0.45\textwidth]{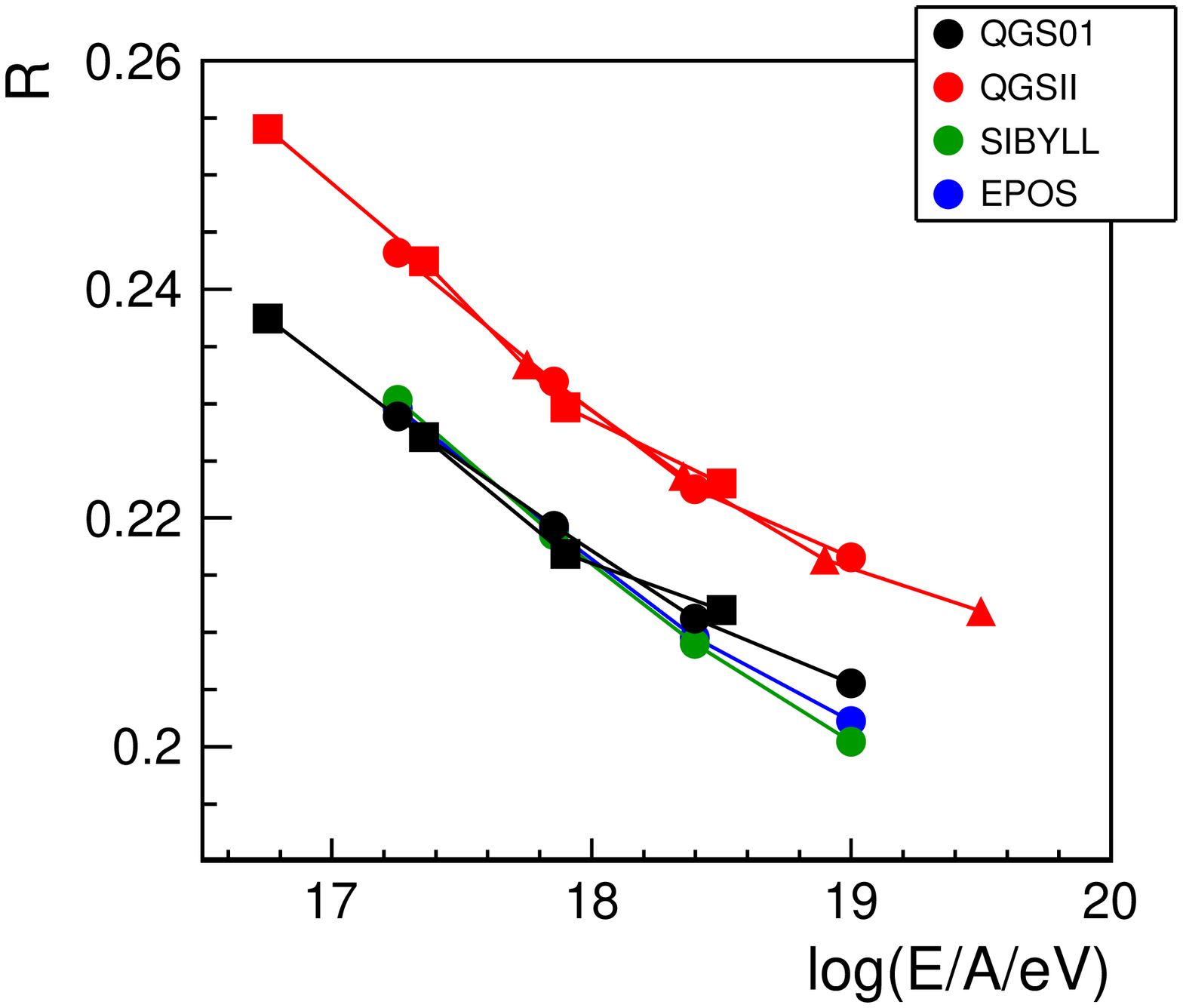}
\caption{Shape parameters {\bf L} (left) and {\bf R} (right)  as a function of energy for four different primaries (p, He, N and Fe) at $\log(E/{\rm eV})=19.0$ in all models, shown in colored dots, and also $\log(E/{\rm eV})=18.5$ in QGSJet-II and QGSJet01c, shown as squares and  $\log(E/{\rm eV})=18.5$ in QGSJet-II, as triangles. In the case of {\bf R}, the energy is scaled by mass number, to highlight the hidden dependency (see text).}
\label{fig:scaling}
\end{center}
\end{figure}

Figure \ref{fig:scaling} shows the evolution of the parameters as a function of energy, for the different models.
Notice that the electromagnetic energy is given by $E_{em}=\sqrt{2\pi} L \left.\frac{dE}{dX}\right|_{max}$ and {\bf L} should be mainly determined by the primary energy. There is a relative change of the order of 5\% between primaries, and of the order of 1\% between models. On the other hand, {\bf R} migth be related to multiplicity, with large differences between primaries. We highlight these trends by using the scaled energy, $\log(E/A)$, inspired in  the superposition model for showers initiated by nuclei. The scaling is reflected within each model, with all primaries and energies lying approximately in two lines: one for QGSJet-II and another one grouping all other models.

\section{Summary and Conclusions}

A Universal Shower Profile can be constructed with a limited number of events to characterize the average behavior of high energy cosmic rays. By constructing an average longitudinal shower profile shape, the statistical errors associated to the detection of single events are significantly reduced. The average shape can be described with the variables {\bf L} and {\bf R}, which contain important information on composition.

While being independent of the first interaction point by construction, the Universal Shower Profile shape keeps information about the other parameters governing shower development which are different among primaries and models, namely multiplicity. Within each model, {\bf L} and {\bf R} give two measurements of $\left< A \right>$, with a reduced dependence on the primary cross-section. The consistency between both is a powerful test of the modeling of the hadronic interactions.

\section*{Acknowledgments}
We would like to thank all our colleagues in the Auger-LIP group for reading the manuscript.
This work is partially funded by Funda\c{c}\~ao para a Ci\^encia e Tecnologia (CERN/FP123611/2011 and SFRH/BPD/73270/2010), 
and funding of MCTES through POPH-QREN-Tipologia 4.2, Portugal, and European Social Fund.


\bibliographystyle{elsarticle-num}
\bibliography{Bib-hX}

\begin{thebibliography}{10}
\expandafter\ifx\csname url\endcsname\relax
  \def\url#1{\texttt{#1}}\fi
\expandafter\ifx\csname urlprefix\endcsname\relax\def\urlprefix{URL }\fi
\expandafter\ifx\csname href\endcsname\relax
  \def\href#1#2{#2} \def\path#1{#1}\fi

\bibitem{USPV}
S.~Andringa, R.~Concei{\c{c}\~a}o, M.~Pimenta, {Mass composition and
  cross-section from the shape of cosmic ray shower longitudinal profiles},
  Astropart.Phys. 34 (2011) 360--367.
\newblock \href {http://dx.doi.org/10.1016/j.astropartphys.2010.10.002}
  {\path{doi:10.1016/j.astropartphys.2010.10.002}}.

\bibitem{UnivM}
J.~Matthews, R.~Mesler, B.~Becker, M.~Gold, J.~Hague, {A Parametrization of
  Cosmic Ray Shower Profiles Based on Shower Width}, J.Phys.G G37 (2010)
  025202.
\newblock \href {http://arxiv.org/abs/0909.4014} {\path{; arXiv:0909.4014}},
  \href {http://dx.doi.org/10.1088/0954-3899/37/2/025202}
  {\path{doi:10.1088/0954-3899/37/2/025202}}.

\bibitem{UnivG}
M.~Giller, A.~Kacperczyk, J.~Malinowski, W.~Tkaczyk, G.~Wieczorek, {Similarity
  of extensive air showers with respect to the shower age}, J.Phys.G G31 (2005)
  947--958.
\newblock \href {http://dx.doi.org/10.1088/0954-3899/31/8/023}
  {\path{doi:10.1088/0954-3899/31/8/023}}.

\bibitem{USPmu}
S.~Andringa, L.~Cazon, R.~Concei\c{c}\~ao, M.~Pimenta, {The Muonic longitudinal
  shower profiles at production}, Astropart.Phys. 35 (2012) 821--827.
\newblock \href {http://arxiv.org/abs/1111.1424} {\path{; arXiv:1111.1424}},
  \href {http://dx.doi.org/10.1016/j.astropartphys.2012.03.010}
  {\path{doi:10.1016/j.astropartphys.2012.03.010}}.

\bibitem{Conex1}
T.~Pierog, et~al., {First results of fast one-dimensional hybrid simulation of
  EAS using CONEX}, Nucl. Phys. Proc. Suppl. 151 (2006) 159--162.
\newblock \href {http://arxiv.org/abs/astro-ph/0411260} {\path{;
  arXiv:astro-ph/0411260}}.

\bibitem{Conex2}
T.~Bergmann, et~al., One-dimensional hybrid approach to extensive air shower
  simulation, Astropart. Phys. 26 (2007) 420--432.
\newblock \href {http://arxiv.org/abs/astro-ph/0606564} {\path{;
  arXiv:astro-ph/0606564}}.

\bibitem{QGSII1}
S.~Ostapchenko, {Non-linear screening effects in high energy hadronic
  interactions}, Phys.Rev. D74 (2006) 014026.
\newblock \href {http://arxiv.org/abs/hep-ph/0505259} {\path{;
  arXiv:hep-ph/0505259}}, \href {http://dx.doi.org/10.1103/PhysRevD.74.014026}
  {\path{doi:10.1103/PhysRevD.74.014026}}.

\bibitem{QGSII2}
S.~Ostapchenko, {On the re-summation of enhanced Pomeron diagrams}, Phys.Lett.
  B636 (2006) 40--45.
\newblock \href {http://arxiv.org/abs/hep-ph/0602139} {\path{;
  arXiv:hep-ph/0602139}}, \href
  {http://dx.doi.org/10.1016/j.physletb.2006.03.026}
  {\path{doi:10.1016/j.physletb.2006.03.026}}.

\bibitem{xsec}
R.~Conceicao, J.~D. de~Deus, M.~Pimenta, {Proton-proton cross-sections: the
  interplay between density and radius}, Nucl.Phys. A888 (2012) 58--66.
\newblock \href {http://arxiv.org/abs/1107.0912} {\path{; arXiv:1107.0912}},
  \href {http://dx.doi.org/10.1016/j.nuclphysa.2012.02.019}
  {\path{doi:10.1016/j.nuclphysa.2012.02.019}}.

\bibitem{QGS01}
N.~N. Kalmykov, Ostapchenko, {Quark-gluon string model and EAS simulation
  problems at ultra-high energies}, Nucl. Phys. Proc. Suppl 52 (1997) 17.

\bibitem{SYBILL21}
E.-J. Ahn, R.~Engel, T.~K. Gaisser, P.~Lipari, T.~Stanev, {Cosmic ray
  interaction event generator SIBYLL 2.1}, Phys.Rev. D80 (2009) 094003.
\newblock \href {http://arxiv.org/abs/0906.4113} {\path{; arXiv:0906.4113}},
  \href {http://dx.doi.org/10.1103/PhysRevD.80.094003}
  {\path{doi:10.1103/PhysRevD.80.094003}}.

\bibitem{EPOS}
K.~Werner, F.-M. Liu, T.~Pierog, {Parton ladder splitting and the rapidity
  dependence of transverse momentum spectra in deuteron-gold collisions at
  RHIC}, Phys.Rev. C74 (2006) 044902.
\newblock \href {http://arxiv.org/abs/hep-ph/0506232} {\path{;
  arXiv:hep-ph/0506232}}, \href {http://dx.doi.org/10.1103/PhysRevC.74.044902}
  {\path{doi:10.1103/PhysRevC.74.044902}}.

\end{thebibliography}

\end{document}